\documentstyle[sprocl,psfig,here]{article}

\def\be{\begin{equation}}
\def\ee{\end{equation}}
\def\bea{\begin{eqnarray}}
\def\eea{\end{eqnarray}}

\begin{document}

\title{Total Cross Sections of ${\bf e^+{}e^-\to hadrons}$ 
and pQCD Tests\footnote{{\small Presented by one of us (K.K.) at the 6th Workshop on Non-perturbative QCD, 
American University of Paris, 5-9 June 2001. Report No. BROWN-HET-1097}}}

\author{K. Kang$^{\rm b,c,d}$, V. V. Ezhela$^{\rm e}$, S. K. Kang$^{\rm d}$,\\
S. B. Lugovsky$^{\rm e}$, M. R. Whalley$^{\rm f}$, O. V. Zenin$^{\rm e}$ \\
(COMPETE $(e^+{}e^-)$ Collaboration) }

\address{$^{\rm b}$Physics Department, Brown University, Providence, R.I. 02912, U.S.A. \\
$^{\rm c}$Institute of Physics {\&} Applied Physics, Yonsei University, Seoul 120-749, Korea \\
$^{\rm d}$School of Physics, Korea Institute for Advanced Study, Seoul 130-012, Korea \\
$^{\rm e}$COMPAS Group, IHEP, Protvino, Russia \\
$^{\rm f}$HEPDATA Group, Durham University, Durham, UK}

\maketitle\abstracts{In the light of recent muon $(g - 2)$ result 
by the E821 experiment at BNL, the importance of and interest in 
the total hadronic cross sections in $e^+{}e^-$ collisions and 
$R(e^+{}e^-\to {hadrons})$ have been heightened. We report 
on the integrated data compilation of available data
of  $e^+{}e^-\to {hadrons}$, transformed to a unique
and meaningful style that implements the pure QED radiative corrections, 
from the PPDS and HEPDATA databases by the COMPETE Group. In particular, 
the continuum data sets are extracted where the direct comparison of the 
parton models and different variants of the pQCD corrections can be made 
and which will be useful for the future refinements of 
$\alpha_{s}(Q^2)$ and $\alpha_{QED}(Q^2)$. 
}

\section{Introduction}

In spite of the tremendous experimental and phenomenological 
efforts that have been devoted to the study of the hadronic production in $e^+ e^-$ collisions, 
it is simply the state of arts that there is no truly integrated 
view on the experimental situation for the main observable such as the total cross section 
for the reaction $e^+{}e^-\to hadrons$ and the ratio
$$ R ={{ \sigma(e^+ e^- \to hadrons)} \over 
{\sigma(e^+ e^- \to \mu^+ \mu^-)_{QED}}}$$
To be specific, we note that some experiments published only the 
data of the cross sections, $\sigma$, while others did only the $R$ data.
The situation is more complicated as for the 
QED radiative effects to the data, because various data sets 
have been corrected by different ways and to different degrees.
For example, in constructing $R$,
two different forms of theoretical calculations had been used, i.e., 
one with $\alpha_{QED}(0)$ and another taking into account the accepted 
form of evolution of $\alpha_{QED}(Q^2)$ then.
Furthermore, some data were never published in numerical form 
and have been read by compilers from graphs.

Determination of the precision data for $\sigma (e^+ e^- \to hadrons)$ and $R$  
has been a continuing endeavor in particle physics as new precision 
experiments became feasible and has received renewed interests recently. 
Some of the recent experimental activities that revived the interest in 
the data of $\sigma (e^+ e^- \to hadrons)$ and $R$ are: the precision 
electroweak data that allowed us to calculate the radiative corrections 
and extract the QED coupling constant; the anomalous magnetic moment of 
the muon  $(g - 2)$ which showed some $2.6 \sigma$ deviation from the 
standard model prediction initially compared to the new result of the 
E281 experiment at BNL\cite{E281}; and the $\tau $ decay data. In the 
analysis of all these experiments, the hadronic contribution is the major 
source of the ambiguity and errors. A new controversy of the the light-by-light 
scattering contribution (as for the sign)\cite{knecht} to the vacuum polarization, 
the dominant leading contribution to the hadronic contribution reinforces only 
the need of precise values of these data.
  
The aim of this work\cite{compete} is to prepare a comprehensive 
compilation of $\sigma$ and $R$ transformed, whenever possible,  
to a unique, and meaningful, style by implementing the pure 
QED radiative corrections. 
The second aim is to construct a procedure to extract 
``continuum data sets" where the direct comparison of 
the parton model with different variants of the pQCD 
corrections can be made. In selecting data for the compilation 
we use the expert assessments from various recent reviews and 
works dealing with precise estimation of running 
$\alpha_{QED}(Q^2)$ and 
$\alpha_{s}(Q^2)$ (see, e.g., Ref.\cite{Eidelman:1995ny}, 
Ref.\cite{Hinchliffe:2000yq}, and Ref.\cite{Zhao:2000xc}).

\section{Data Selection and Normalizations} 

All published data were extracted in  numerical form from 
the PPDS RD \cite{PPDS} database and from the
Reaction Database of HEPDATA system \cite{hepdata}. 
Certain data were excluded based on the following exclusion criteria:

\begin{itemize}
\item All preliminary data were excluded;
\item Data obtained under incomplete kinematical conditions,
and not extrapolated to the complete kinematic region, were excluded;
\item Data in the form of dense energy scans are omitted if the authors published
in addition the data averaged over wider energy bins.
\end{itemize}

The data sets that survived the above exclusion criteria were
subdivided into the following six categories:\\

\boldmath
{\large\bf $\sigma_1^{sd}$}: 
\unboldmath
~ Cross section data corrected in the original works
for the contributions of two-photon exchange diagrams,
the initial state bremsstrahlung and for the initial state vertex loops,
but not corrected for the vacuum polarization contribution to
the running $\alpha_{QED}$. These  ``{\bf semi-dressed}" cross sections
take into the full $\gamma/Z$ propagator with  vacuum polarization effects.

\boldmath
{\large\bf $\sigma_2^{sd}$}: 
\unboldmath
~ Cross section data radiatively corrected  according to the procedure
of Bonneau and Martin \cite{Bonneau:1971mk} in the original works.
This procedure included radiative corrections for the
initial state radiation, electronic vertex correction, and
the correction for the electronic loop in the photon propagator.
Thus it partially took into account 
the vacuum polarization effects.
To obtain from these cross sections the ``{\bf semi-dressed}" ones
we rescale
\boldmath
$\sigma_2^{sd}$
\unboldmath 
by a factor
$1/(1-\Delta\alpha_{QED}^e{}(s))^2$.

\boldmath
{\large $R_3^{bare}$}:
\unboldmath
~Data on $R$
obtained from measurements of the ``{\bf semi-dressed}" 
\boldmath
($\sigma^{sd}_2$)
\unboldmath
cross-section divided by the point-like muonic
cross section with fixed $\alpha_{QED} = \alpha_{QED}(0)$ in the works.
To obtain the ``{\bf bare}" $R$-parameter
we rescale these data  by the factor
$(\alpha_{QED}(0)/\alpha_{QED}(s))^2/(1-\Delta\alpha^e_{QED}(s))^2$.

\boldmath
{\large $R_4^{bare}$}: 
\unboldmath
~Data on $R$  obtained from measurements of the ``{\bf semi-dressed}"
\boldmath
($\sigma^{sd}_1$) 
\unboldmath
cross section divided
by the point-like muonic cross section, in the original works,
but with the running $\alpha_{QED}(s)$ which  thus obtained the ``{\bf bare}" 
$R$-ratio.

\boldmath
{\large $\sigma_5^{sd}$}: 
\unboldmath
~ LEP I cross section data at the $Z$ peak not corrected for initial
state radiation and
electronic QED vertex loops, and LEP II -- III cross section data at
$\sqrt{s} > 130$ GeV
with a cut \mbox{$s'/s = 1 - (\sqrt{s}/2)(1 - 0.7225)$},  
$\sqrt{s'}$ being an effective mass of the propagator. 
%
These data were rescaled to the ``{\bf semi-dressed}" cross section 
\boldmath
$\sigma_1^{sd}$
\unboldmath
 as follows.
First, the theoretical cross sections\footnote{Values for
$\sigma^{th}_{cut}$ were calculated using ZFITTER 
subroutine ZUTHSM with argument settings according to the cuts applied
in the LEP I-II-III measurements.
Values for $\sigma^{th}_{born}$ were calculated by the same subroutine but 
with the special flag switching it to calculate the IBA cross sections.}
 ~$\sigma^{th}_{cut}$  and 
$\sigma^{th}_{Born}$ were calculated using ZFITTER 6.30
package ~\cite{Bardin:2001yd}, 
assuming the values of the standard model parameters from PDG~\cite{Groom:2000in}. 
Here $\sigma^{th}_{cut}$ means the cross section measured
with cuts applied in the particular experiment and determined from the fit 
by ZFITTER and $\sigma^{th}_{Born}$ denotes the theoretical cross section,
calculated by ZFITTER in the Improved Born Approximation (IBA), 
that includes the full $\gamma / Z$ propagator with vacuum polarization effects.
Then the experimental cross section was multiplied by
the factor $\sigma^{th}_{Born}/\sigma^{th}_{cut}$.

\boldmath
{\large $\sigma_6^{sd}$}: 
\unboldmath
~Low energy data ($2m_\pi < E_{cm} < 2$ GeV), the treatment of which  requires special consideration.

All the cross sections $\sigma{}(e^+{}e^-\to hadrons)$
in this range are obtained by summing the  exclusive channels
and therefore give only a lower estimate of
the total hadronic cross section.
Below 1 GeV we summed up the $2\pi$ and $3\pi$ channels
using a  linear
interpolation of the individual data sets within specific energy regions and
combined them  to give the total hadron cross sections in these regions.
The errors were calculated
according to these interpolation and summation procedures.

Such an approach works well if all the data are evenly 
distributed over $\sqrt{s}$ and have comparable errors.
Otherwise it may cause false resonance-like structures 
in the total cross section in the given $\sqrt{s}$ interval if
there are few  points of the leading channel with large errors,
and  if this interval is filled more densely by the points of 
the minor channels. 
Such false structures are possible and likely to occur for 
the exclusive channel data in the range $1.4 < \sqrt{s} < 2$ GeV,
where we summed the contributions from the channels 
$\ge 3 hadrons$, $\pi^+\pi^-$, $K^+{}K^-$, and $K_S K_L$.
To avoid such false structure  we have
determined the sum of  exclusive channels
not for all points, where at least one channel is measured,
but in a fewer number of points, for which the channels yielding 
the major contributions are identified.
In other aspects the summation procedure was the same as the one 
used for the data below 1 GeV.
The exclusive sum data in the $0.81 < \sqrt{s} < 1.4$ GeV region
are taken from the paper\cite{Dolinsky:1991vq}.  
The data used for the exclusive channel summation
around $\phi$ resonance ($0.997 < \sqrt{s} < 1.028$ GeV) 
were corrected for the initial state radiation,
electronic vertex loops and leptonic 1-loop insertions
into the $\gamma$ propagator in these data.
We have properly rescaled these data.
In the remaining $0.81$--$1.4$ GeV data they applied 
QED-corrections using the Bonneau and Martin~\cite{Bonneau:1971mk} 
prescription and all these data points were rescaled as in the
{\boldmath $\sigma_2^{sd}$\unboldmath}-case.
\\

\section{Data Compilations on {$\sigma^{sd}$} and {$R^{bare}$} } 

After selecting and rescaling the data, 
normalized to the full $\gamma / Z$ contributions,
we assemble the complete data set of the total
hadronic cross sections {\boldmath $\sigma^{sd}$\unboldmath},
in accordance with the symbolic relation

{\boldmath $$\sigma^{sd} = \sigma_1^{sd} \cup \sigma_2^{sd} \cup \sigma_5^{sd}
\cup \sigma_6^{sd} \cup \left [ ( R_3^{bare} \cup R_4^{bare}) \cdot
\sigma^{\mu\mu}_{QED~pole,~running~\alpha_{QED}}\right ]$$ \unboldmath}
and the complete data set for the $R$-ratios\\
{\boldmath $$ R^{bare} ={{ \sigma(e^+ e^- \to hadrons)} \over 
{\sigma(e^+ e^- \to \mu^+ \mu^-)_{QED~pole,~running~\alpha_{QED}}}}.$$ \unboldmath}

As described above, in creating compilations which are uniform in the sense of
implementing a standardized pure QED radiative corrections to
the ``raw" published experimental data, the values of the running 
$\alpha_{QED}(s)$ and $\Delta\alpha_{QED}^{e}(s)$ are needed. 
The numerical values of these parameters and estimates for their uncertainties are described 
in the next section. Both compilations are stored and maintained in the PPDS CS 
database, corresponding to the record of the 
original data but with the rescaled data points. Also stored are
brief descriptions of the applied conversion procedures 
in the special comment in each record.
The {\boldmath $\sigma$\unboldmath} and {\boldmath $R^{bare}$\unboldmath} 
compilations obtained by this procedure are shown
in the Figs.~1 and 2, respectively.

\section{$\alpha_{QED}$, $\Delta \alpha^{had}_{QED}$ and the Continuum Region}

\subsection{Theoretical Relations.}
 
The QED running coupling constant
$\alpha_{QED}(s)$ can be expressed in the form 

\begin{eqnarray}\label{alpha_QED}
\alpha_{QED}(s) &=& \frac{\alpha(0)}{1 - \Delta \alpha (s)}~,
\nonumber\\
\Delta \alpha (s) &=& \Delta \alpha^{had} (s) + \Delta \alpha^{lep} (s)~,
\end{eqnarray}
where $\Delta \alpha^{had}(s)$, $\Delta \alpha^{lep} (s)$
are hadronic and leptonic contributions to the QED vacuum polarization,
respectively.

$\Delta \alpha^{lep} (s)$ is well approximated in perturbative QED
by the sum of leptonic loop contributions,
%
\begin{equation}\label{Delta_alpha_lep}
\Delta\alpha^{lep} (s) = \sum_{l=e,\mu,\tau} \frac{\alpha(0)}{3\pi}
\left[ \ln\frac{s}{m_l^2} - \frac{5}{3} + {\cal O}\left(\frac{m_l^2}{s}\right)\right]~.
\end{equation}
The situation with respect to $\Delta \alpha^{had} (s)$ is more complicated 
due to an essentially non-perturbative 
nature of the strong interaction
at low energies.  Using the unitarity and the analyticity of the
scattering amplitudes, one can express $\Delta \alpha^{had} (s)$
in the form of a subtracted dispersion relation
%
\begin{equation}\label{disp_rel}
\Delta\alpha^{had} (s) = - \frac{\alpha(0) s}{3\pi}
\int_{4m_\pi^2}^{\infty} \frac{R(s') ds'} {s' (s' - s - i0)}~,
\end{equation}
where $R(s)$ is the ``bare'' hadronic $R$-ratio.

This relation allows to effectively evaluate
the $pQCD$ $R$-ratios in the continuum $\sqrt{s}$ intervals 
in terms of experimental $R$ data containing the non-perturbative effects.

We have evaluated the dispersion integral in two parts, 
\begin{equation}\label{disp_rel_comb}
\Delta\alpha^{had} (s) = - \frac{\alpha(0) s}{3\pi}
\left [ 
\int_{4m_\pi^2}^{s_{cut}} \hspace*{-2.5ex} ds'~\frac{R^{data}_{bare}(s')} {s' 
(s' - s - i0)} +
\int_{s_{cut}}^{\infty} \hspace*{-2ex} ds'~\frac{3\sum_q Q^2_q (1+\frac{\alpha_S(s')}{\pi})} 
{s' (s' - s - i0)} \right ]~,
\end{equation}
where the first integral was calculated numerically as a trapezoidal sum over 
the weighted average of experimental $R$ points in the range 
$2m_\pi < \sqrt{s'} < \sqrt{s_{cut}} = 19.5$ GeV\footnote{Choosing 
$s_{cut}$ we must satisfy at least the following 
two requirements:\\
1)~ $s_{cut}$ should lie well above all the hadronic resonances, i.e.
1-loop perturbative QED can be used to calculate hadronic vacuum
polarization at $s > s_{cut}$;\\
2)~ there should be enough densely distributed experimental points at 
$s < s_{cut}$ for the trapezoidal 
numerical evaluation of the dispersion 
integral over $2m_\pi < \sqrt{s} < \sqrt{s_{cut}}$.
As there are few experimental points in the interval
$13 < \sqrt{s} < 30$ GeV, $\sqrt{s_{cut}} = 19.5$ GeV
appears to be a compromise between these requirements.}

Our numerical evaluation procedure is in general similar to the one 
applied in Ref.~\cite{Eidelman:1995ny}.\footnote{As the calculation of
$R^{data}_{bare}$ from the original $\sigma$ and $R$ data
(see Section 1) in turn requires the evaluation
of $\Delta\alpha^{had}(s)$, we applied the following simple
iterative procedure: 
Taking as the zeroth approximation $R(s)$ as obtained from the original data
using  1-loop perturbative $\alpha_{QED}(s)$,
we evaluate $\Delta\alpha^{had(1)}(s)$, which in turn is used
to compute the  next approximation
of $R(s)$ from the original data, and then 
evaluate the integral again to obtain $\Delta \alpha^{had(2)}(s)$, 
and so on.
This process converges well and after the two iterations 
the algorithmic error of $\Delta\alpha^{had}(s)$
in the continuum region 
is much less than the error coming from 
the experimental data uncertainty.}

We obtained $\Delta \alpha^{had}(M_Z^2) = 0.0274\pm 0.0004\mbox{(exp.)}$,
which is consistent with the results
$0.027382 \pm 0.000197$ and $0.027612 \pm 0.000220$, obtained with two
methodically different low energy data sets in Ref.\ \cite{Martin:2000by}.
This result is preliminary and serves only as a guarantee against 
egregious errors in our procedure.
The procedure of $\Delta \alpha^{had}(M_Z^2)$ evaluation requires
further refinement.


~In the parton model, before QCD corrections are applied,
 the $R$-ratio is given by
\begin{equation}\label{R0}
R = 3 \sum_q R^0_q = 
3 \sum_q \left [ \beta_q (1 + \frac{1}{2}(1-\beta_q^2))\cdot R_q^{VV} 
+ \beta_q^3 R_q^{AA}\right ]~,
\end{equation}
where
\begin{eqnarray}\label{RVV_RAA}
R_q^{VV} &=&
e_e^2 e_q^2 + 2e_e e_q \overline{v}_e\overline{v}_q~ \mbox{Re}\chi 
+ (\overline{v}_e^2 + \overline{a}_e^2)\overline{v}_q^2~ \vert\chi\vert^2
\nonumber\\
R_q^{AA} &=&
(\overline{v}_e^2 + \overline{a}_e^2)\overline{a}_q^2~ \vert\chi\vert^2
\end{eqnarray}
\noindent
with
\begin{eqnarray}
\chi &=& \frac{1}{16\overline{s}^2\overline{c}^2}
\frac{s}{(s - M_Z^2 + i M_Z\Gamma_Z)}
\nonumber\\
\overline{v} &=& \sqrt{\rho}~ (T_3 - 2Q\overline{s}^2)
\nonumber\\
\overline{a} &=& \sqrt{\rho}~ T_3
\nonumber\\
\rho &=& 1 + \frac{3\sqrt{2}G_F}{16\pi^2} m_t^2 + \cdots~~.
\end{eqnarray}
Here $\overline{s}^2, \overline{c}^2$ are  effective $\sin^2\theta_W, 
\cos^2\theta_W$  defined through renormalized couplings at $s = M_Z^2$ and
\begin{equation}
M_Z^2 = \frac{\pi \cdot \alpha_{QED}(M_Z^2)}
{\sqrt{2}G_F \cdot \rho \cdot \overline{s}^2 \cdot \overline{c}^2}~.
\end{equation}

The dominant correction to $\rho$ originates from $t$-quark loops
in $W$ and $Z$ propagator due to
the large mass splitting between the $b$ 
and $t$ quarks 
which represents an $SU(2)$ breaking in the unified electroweak theory.

Including QCD loops, $R$ is given by (see, e.g., Ref.\cite{Chetyrkin:1996ia} and
Ref.\cite{Chetyrkin:2000zk}):

\begin{equation}
\begin{array} {rcl}
R &=& 3\sum R_q \\
  &=& 3\sum \left[ R_q^0  +  \left( {\displaystyle \frac{\alpha_s} {\pi}}
      + \left( {\displaystyle
      \frac{\alpha_s} {\pi}} \right)^2 (1.9857-0.1153 N_f) \right. \right. 
\nonumber \\
  &-& \left. \left. \left( {\displaystyle \frac{\alpha_s} {\pi}} \right)^3 
 (6.6368 + 1.2001 N_f + 0.0052 N_f^2 + 1.2395 (\sum Q_q)^2 ) \right) \right. 
\nonumber \\
&\times& \left. \left. \left. (f_1 R_q^{VV} + f_2 R_q^{AA}) 
         \right. \right. \right ] 
\end{array}
\label{R}
\end{equation}
where $N_f$ is the number of active quark flavors.
\noindent
The coefficients $f_1$ and $f_2$ depend only on the quark velocity 
$\beta = (1- 4M_q^2/s)^{1/2}$, where $M_q$ is the threshold for the flavor production
of quark $q$.
Schwinger~\cite{Schwinger:1970??} calculated $f_1$ in the QED context, which is accurate enough for our purposes:

\begin{equation}
f_1 = \beta_q \left ( 1 + \frac{1}{2}(1-\beta_q^2) \right ) 
\frac{4\pi}{3}\left ( \frac{\pi}{2\beta_q} - \frac{3+\beta_q}{4} 
\big{(}\frac{\pi}{2} - \frac{3}{4\pi}\big{)}\right )
\end{equation} 

The coefficient $f_2$ (which have no counterpart in QED) were
calculated by Jersak et al~\cite{Jersak:1981uv}, which 
we parameterized as 
\begin{equation}\label{f2}
f_2 = a_4\beta_q^4 + a_3\beta_q^3 + (1 - a_4 - a_3) \beta_q^2
\end{equation}
with $a_4 = -16$, $a_3 = 17$.
From equations~(\ref{R0}), (\ref{RVV_RAA}), (\ref{R}) one can easily
see that the relative error of $R$ is less than $10^{-7}$ at $\sqrt{s} = 10$ GeV
and less than 0.5\% at the $Z$ pole, even if we allow a 100\% error for $f_2$. 
As we did not use $Z$ pole data in our fits,
this could not be a source of large theoretical errors.  

In the massless quark limit, $f_1 = f_2 = 1$.
However, at $\sqrt{s} = 35$ GeV for $b$ quark one has $\beta = 0.963$
and thus $f_1\simeq 1.3$ and $f_2\simeq 1.7$.
Indeed, such a parameterization of mass effects 
is valid in QCD only in ${\cal O}(\alpha_s)$ order. 
Now the correct parameterization is known up to
${\cal O}(\alpha_s^3)$ order 
(see, e.g., Ref.\cite{Chetyrkin:1996ia} and Ref.\cite{Chetyrkin:2000zk}),
but it has not been implemented yet into our programme. 
It may be that it results in large enough 
discrepancies just in the region where $\alpha_s/\pi$ becomes large. 

We used the following three-loop parameterization for $\alpha_s$ 
(see, e.g., Ref.\cite{Marshall:1989ri}):
\begin{eqnarray}
\ln\big{(}Q^2/\Lambda^2_{\overline{MS}}\big{)} &=& 
\frac{4\pi}{\beta_0\alpha_s}~
- ~\frac{1}{2}\frac{\beta_1}{\beta_0^2}
\ln\left [ \left ( \frac{4\pi}{\beta_0\alpha_s}\right)^2 
          + \frac{\beta_1}{\beta_0^2}
            \left ( \frac{4\pi}{\beta_0\alpha_s}\right )
          + \frac{\beta_2}{\beta_0^3}\right ]
\nonumber\\
&& \phantom{\frac{4\pi}{\beta_0\alpha_s}~}
 - ~\frac{1}{\Delta}\left (\frac{\beta_1}{\beta_0^2}\right )
    \tan^{-1} \left [ \frac{1}{\Delta} \left ( \frac{\beta_1}{\beta_0^2}
   + \frac{2\beta_2}{\beta_0^3}
      \left ( \frac{\beta_0\alpha_s}{4\pi}\right ) \right ) \right ]
\end{eqnarray}
with
\begin{eqnarray}
\beta_0 &=& 11 - 2N_f/3~,
\nonumber\\ 
\beta_1 &=& 2(51 - 19N_f/3)~,
\nonumber\\ 
\beta_2 &=& (2857 - 5033N_f/9 + 325N_f^2/27)/2~, 
\nonumber\\ 
\Delta  &=& \sqrt{4\beta_2/\beta_0^3 - \beta_1^2/\beta_0^4} ~.
\nonumber\\ 
\end{eqnarray}
The onset of a new flavor $q$ 
takes place at $Q^2 = \mu^2_q \equiv 4 m_q^2$, where, in general, $m_q\neq M_q$.
There are five 
different $\Lambda_{\overline{MS}}(N_f)$ corresponding to $N_f = 2,3,4,5,6$
above the appropriate quark thresholds.
To sew up $\alpha_s$ at the thresholds we need to use a matching condition 
for $\Lambda$'s: $\alpha_s(Q^2)$ is determined by fixing, say, $\alpha_s(M_Z^2)$, 
which is the parameter determined from the fits in our work.

We solved this equation at given $Q^2$ numerically by rewriting it
in the form $x = f(x)$, where $x = 4\pi/(\beta_0\alpha_s)$.
It enabled us to obtain correct $\alpha_s$ even at $Q^2/\Lambda^2\sim 1$, 
where the expansion by powers of leading logarithms 
is no more valid (see, e.g., Ref.\cite{Hinchliffe:2000yq}).
At $Q^2/\Lambda^2\gg 1$ both methods yield the same result
(see Fig.~4).

\subsection{Preliminary Fit Results}
In order to check the consistency of the description
of the data on
$$R = \sigma(e^+{}e^-\to hadrons)/\sigma(e^+{}e^-\to\mu^+\mu^-)$$
with the theoretical $R$-ratio expression
we have performed several fits using a sub-set of the compiled
data set of $R$ which is applicable to the perturbative domain.
This sub-set of the data is defined by the the following steps:
\begin {enumerate}
\item  The low energy region $2m_\pi < \sqrt{s} < 2$ GeV is completely excluded;
\item All the data above 70 GeV are excluded, as we did not attempt
to perform any fits of $Z$ pole parameters $M_Z$, $\Gamma_Z$,
$\overline{s}^2_W$ and $M_{top}$. This exclusion is justified because the
data above 70 GeV 
weakly depend on $\alpha_s(M_Z)$ and the quark masses and thus their 
influence on the fit of $Z$ pole parameters would be rather negligible;
\item Data covering the narrow hadronic $1^{--}$ resonances, 
$J/\psi(1S)$, $\psi(2S)$, $\psi(3770)$,
$\Upsilon(1S)$, $\Upsilon(2S)$, $\Upsilon(3S)$, $\Upsilon(4S)$,
$\Upsilon(10860)$ and $\Upsilon(11020)$ are completely excluded.
(The visible widths of these resonances are determined by the 
machine energy spread $\Delta E_{beam} \sim 10$~MeV.);
\item Also completely excluded are the SLAC-SPEAR-MARK-1 
data
in the $2.6$ GeV $< \sqrt{s} < 7.8$ GeV range.
These data systematically lie $\simeq$15\% above 
other experiments in the same interval.
\end{enumerate}

%
We distinguished the $\gamma^*\to q\overline{q}$ 
production threshold quark masses $M_q$
from the QCD quark ``masses'' $m_q$ determining 
the energy scale of the onset of the new flavour $q$ 
in the running of $\alpha_S(s)$.
%
After performing the exclusions 1) -- 4) 
we fixed $\alpha_s(M_Z)$, $M_Z$, $\Gamma_Z$, $\overline{s}^2_W$,
$M_q$ ($q = u,d,s,t$), $m_q$ ($q = u,d,s,c,b,t$)
at the values shown in the Table~1, leaving just $M_c$ and $M_b$ as 
free parameters.
Another essential feature of our fit ({\bf I})
was that we retained as free parameters the
left and right boundaries of the interval $[\sqrt{s_1},\sqrt{s_2}]$,
$3.0\le\sqrt{s_1}\le 3.670$ GeV, $3.870\le\sqrt{s_2}\le 5.0$ GeV,
which was to be excluded from the 
data set in the process of $\chi^2/dof$ minimization itself.\footnote{Narrow
$\psi$ family resonances between 3 and 4 GeV were excluded by default.
Somewhat controversial is the question, whether the gaps between $\psi$'s
can be really treated as the continuum regions.
$R$ ratio demonstrates no step up until $\sqrt{s} = 3.9-4.0$ GeV, just at the 
left knee of the broad $\psi(4040)$ resonance, where the continuum
approximation of $R$ does not work. 
Nevertheless, this step results in the fitted $c$-quark mass
$M_c = 1.971$ (see fit ({\bf III}) in the Table~1)}.
Thus, the number of the degrees of freedom was also variable
during this fit, for which the standard MINUIT\cite{James:1975dr} program is used.
Such a $\chi^2/dof$ minimization procedure
cancelled possible arbitrariness in the exclusions
of broad resonances in the region above $c\overline{c}$ threshold.

In the fit ({\bf II}) we released $\alpha_s(M_Z)$ free, fixing at the same time 
the excluded region $\sqrt{s} = 3.09 \div 4.44$ GeV.
$M_c$, $M_b$ were fixed at their fitted values obtained in the fit ({\bf I}). 
Fit ({\bf II}) gives somewhat high value $\alpha_s(M_Z) = 0.128\pm 0.032$.

\newpage

\begin{table}[htbp]
\paragraph{Table 1.} Fit results in the defined ``continuum" regions. 
Adjustable parameters are shown in bold.
\vspace{0.5cm}

\begin{center}
\begin{tabular}{|c|c|c|c|}
\hline
&&&\\[-2mm]
Parameter&{\bf I}& {\bf II} & {\bf III} \\[2mm]
\hline
$\alpha_s(M_Z^2)$ & 0.1181 
                  & {\bf 0.128(32) }  
                  &{\bf 0.126(37)}
\\
\hline
$M_Z$ & 91.187 & 91.187 & 91.187\\
\hline
$\Gamma_Z$ & 2.4944 & 2.4944 & 2.4944\\
\hline
$\overline{s}^2$ & 0.23117 & 0.23117 & 0.23117\\
\hline
$M_t$ & 174.3 & 174.3 & 174.3\\
\hline
$M_u$ & 0.140 & 0.140 & 0.140\\
$M_d$ & 0.140 & 0.140 & 0.140\\
$M_s$ & 0.492 & 0.492 & 0.492\\
\hline
{\bf $M_c$} & {\bf 1.500(18) } 
            & 1.5 
            & {\bf 1.9710(1)} 
\\ 
{\bf $M_b$} & {\bf 5.23   $\pm$ 1.16 } 
            & 5.23
            & {\bf 6.0 $\pm$ 1.4}
\\
\hline
$m_u$ & 0.003 & 0.003&0.003\\
$m_d$ & 0.006 & 0.006 &0.006\\
$m_s$ & 0.120 & 0.120 &0.120\\
$m_c$ & 1.2   & 1.2  & 1.2\\
$m_b$ & 4.2  & 4.2 &4.2\\
$m_t$ & 174.3 & 174.3&174.3\\
\hline \hline
		    &2$m_\pi \div$ 2.0
		    &2$m_\pi \div$ 2.0
		    &2$m_\pi \div$ 2.0
\\	
                    &   3.093$\div$3.113 
		    &	
		    & 3.093$\div$3.113
\\
		    &
                    &
		    & 3.684$\div$3.688
\\
            	    &
		    &
                    & 3.670$\div$3.870
\\ \cline{2-4}
{\bf Excluded}      &   3.175(247) $\div$\phantom{\bf 4.44}
                    &   3.09 $\div$4.44
                    &   4.000$\div$4.400
\\
                    &   {\bf 4.319(105)}   
                    &   
                    &   
\\ \cline{2-4}
   $\sqrt{S}$       &   9.450$\div$9.470 
                    &   9.450$\div$9.470 
                    &  9.450$\div$9.470
\\
                    &   10.000$\div$10.025 
                    &   10.000$\div$10.025
                    & 10.000$\div$10.025
\\
{\bf intervals}     &   10.34 $\div$10.37  
                    &   10.34 $\div$10.37 
                    &  10.34 $\div$10.37
\\
                    &   10.52 $\div$10.64 
                    &  10.52 $\div$10.64
                    &  10.52 $\div$10.64
\\
                    &   10.75 $\div$10.97 
                    &  10.75 $\div$10.97
                    &  10.75 $\div$10.97
\\
                    &   11.00 $\div$11.20 
                    &  11.00 $\div$11.20
                    &  11.00 $\div$11.20
\\
                    &  70 $\div$ 188.7
                    &  70 $\div$ 188.7
                    &  70 $\div$ 188.7
\\
\hline  \hline
&&&\\[-2mm]
{\bf $\chi^2/$dof} & {\bf 0.690} 
                   & {\bf 0.665} 
                   & {\bf 0.822}
\\[2mm] 
\hline
\end{tabular}
\end{center}
\end{table}

\section{Summary. Assessed Data Compilations.}

In summary:
\begin{itemize}
\item 
We have created two complementary computerized ``raw" data 
compilations on $\sigma$ and $R$ with data presented as in the original 
publications in all cases except the low energy subsample. In the
low energy region,
where there are no direct measurements of the total cross 
section, we obtain estimates of the total cross section either as the 
sum of exclusive channels or as the sum of the two-body exclusive 
channels with the data on $e^+~e^- \to \ge 3 ~ hadrons$.  
\item On the basis of above two data sets we have created
compilations of data on {\boldmath $\sigma^{sd}$\unboldmath}
and
{\boldmath $R^{bare}$\unboldmath} 
 with one-to-one
correspondence between the data points. Where necessary the data have been rescaled to the
standard style of implementing the pure QED radiative corrections to the initial
state and to the photon propagator to produce a data set which is suitable to test  the parton
model with pQCD corrections to {\boldmath $R^{bare}$\unboldmath}, and to be able to obtain more reliable
estimates for $\Delta\alpha_{QED}^{had}(Q^2)$.
\item  From the total data compilation on {\boldmath $R^{bare}$\unboldmath} we have defined a
``continuum" data compilation sub-set which can be used in conjunction with other
data in the refinements of the $\alpha_s(Q^2)$ evolution and possibly in 
the global fits of the Standard Model.
\end{itemize}

All data files are accessible by: 
\begin{center}
{\rm http://wwwppds.ihep.su:8001/comphp.html}
\end{center}
and will be accessible from the PDG site soon.
\section*{Acknowledgments}
This work is supported in part by the U.S. DoE Contract DE-FG02-91ER4068-Task A,
the grant RFBR-01-07-90392 and a grant from PPAR(UK). 
One of us (KK) would like to thank Bruce McKellar for discussions 
and also KIAS and Yonsei University, Seoul, Korea for 
the hospitality during the sabbatical year 2000 - 01.

%


\newpage

\section*{Figures}
\begin{figure}[H]
\psfig{file=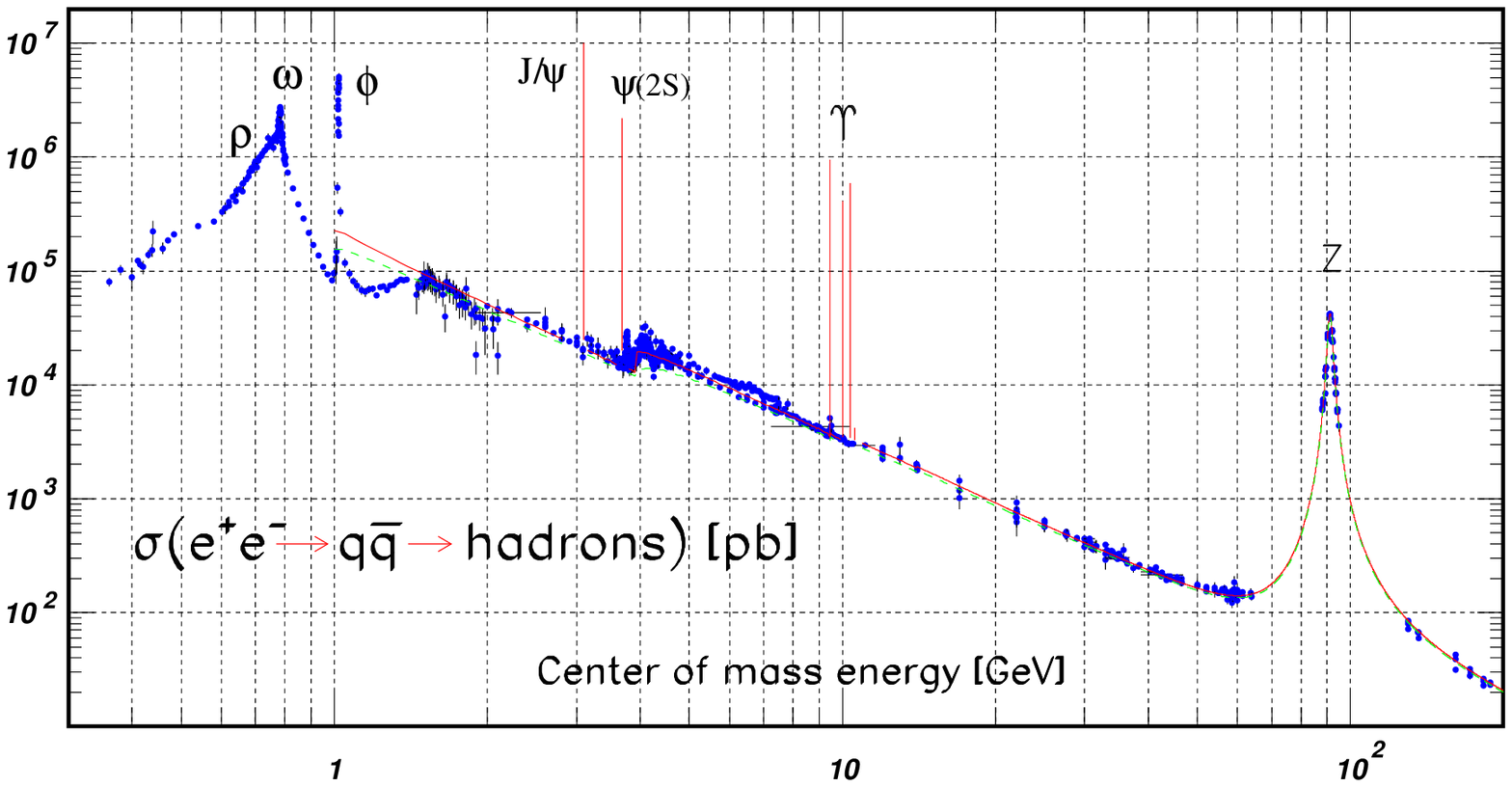,width=140mm,bbllx=50pt,bblly=30pt,bburx=617pt,bbury=283pt}
\caption{World data on the total cross section of
$e^+{}e^-\to hadrons$. 
The experimental data on  $J/\psi$, $\psi(2S)$, and $\Upsilon(nS), n=1..4$ 
resonances is not shown.
Curves are an educational guide.
Solid curve is the prediction of the cross section 
in the three-loop
QCD approximation with non-zero quark masses.
Dashed curve is a ``naive" quark parton model prediction.}
\end{figure}
%
%
\begin{figure}[H]
\psfig{file=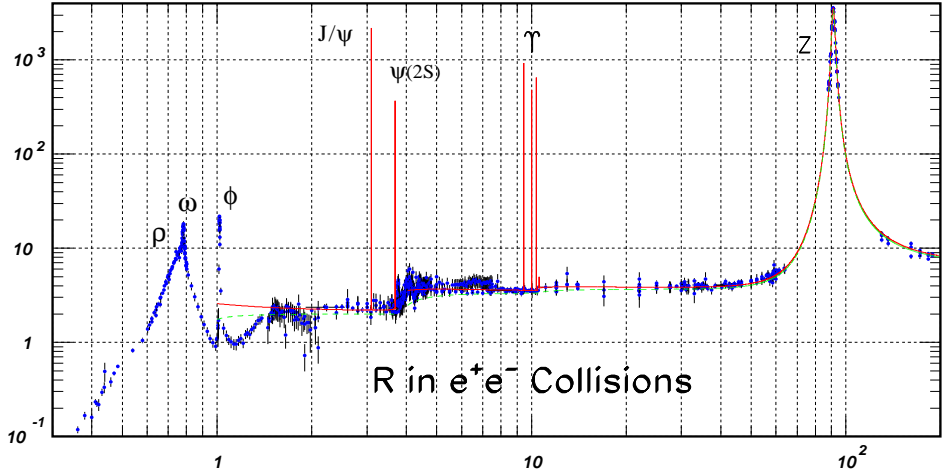,width=140mm,bbllx=50pt,bblly=35pt,bburx=617pt,bbury=273pt}
\caption{$R$-ratio. 
Data set is the same as in Figure~1. 
Solid curve is the $R$-ratio prediction in the three-loop
QCD approximation with non-zero quark masses.
Dashed curve is a ``naive" quark parton model prediction
for the ratio parameter $R$.
}
\end{figure}
\begin{figure}[H]
\psfig{file=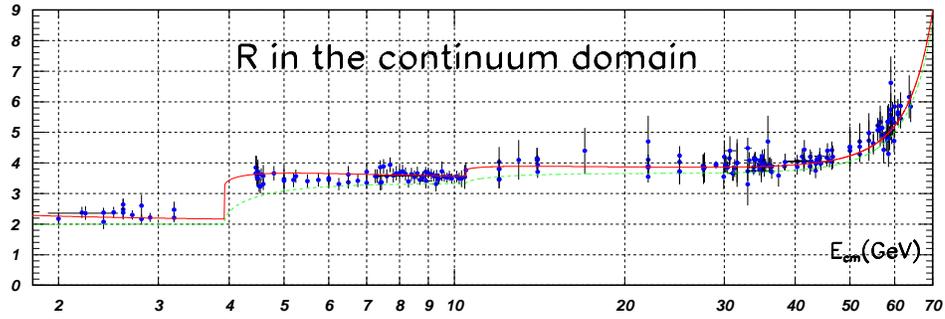,width=135mm,bbllx=50pt,bblly=10pt,bburx=910pt,bbury=283pt}
\caption{Data set used for the preliminary fits. Curves are the same as in Figure~2.}
\end{figure}
\begin{figure}[H]
\psfig{file=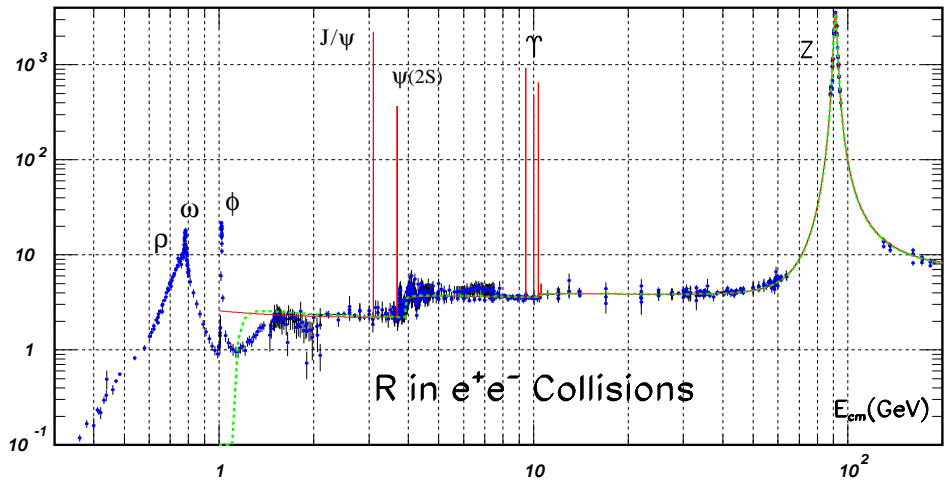,width=140mm,bbllx=50pt,bblly=30pt,bburx=617pt,bbury=283pt}
\caption{Comparison of $R$ parameters
obtained with $\alpha_s(s)$, evaluated by our numerical method (solid curve)
and by the method described in 
[5] (dashed curve).}
\end{figure}

\end{document}